\documentclass[reqno]{amsart}
\usepackage{a4wide}

\numberwithin{equation}{section}

\def\R{{\mathcal R}}
\def\S{{\mathcal S}}

\def\D{{\mathcal D}}

\newtheorem{lemma}{Lemma}

\newtheorem{conjecture}{Conjecture}

\begin{document}

\title[Deformations and approximately integrable systems]
{Deformations of the Monge/Riemann hierarchy and approximately integrable systems}

\author{I.A.B. Strachan}
\date{\today}
\address{Department of Mathematics\\ University of Hull\\
Hull HU6 7RX\\ U.K.}

\email{i.a.strachan@hull.ac.uk}

\begin{abstract}
Dispersive deformations of the Monge equation $u_u=uu_x$ are studied using
ideas originating from topological quantum field theory and the deformation
quantization programme. It is shown that, to a high-order, the symmetries of
the Monge equation may also be appropriately deformed, and that, if they exist
at all orders, they are uniquely determined by the original deformation.
This leads to either a new class of integrable systems or to a rigorous notion of
an approximate integrable system. Quasi-Miura transformations are also constructed
for such deformed equations.
\end{abstract}

\maketitle

%

\bigskip

\section{Introduction}

Consider a general scalar evolution equation
\begin{equation}
u_t = K(u)
\label{basicexample}
\end{equation}
where $K(u)$ is a smooth function depending on $u$ and the
$x$-derivatives of $u$, but not explicitly on the independent
variable $x\,.$ A characteristic feature of an integrable
system is the existence of an infinite hierarchy of compatible,
commuting flows, these being the generalized symmetries of the original equation
(\ref{basicequation}). Such symmetries have the form
\[
u_\tau =Q(u)
\]
(where again $Q$ is a smooth function of $u$ and the
$x$-derivatives of $u$, but not explicitly on the independent
variable $x$) and must satisfy the equation
\[
K^\prime Q - Q^\prime K = 0
\]
where the prime denotes the Fr\'echet derivative
\[
K^\prime(u) Q = \left.\frac{\partial~}{\partial \epsilon} K(u+\epsilon Q)
\right|_{\epsilon=0}\,.
\]
The paradigm for such constructions is the Korteweg-deVries equation
\begin{equation}
u_t = u u_x + u_{xxx}\,.
\label{KdV}
\end{equation}
For this equations, the hierarchy of symmetries may be constructed by
exploiting the bi-Hamiltonian structure of the KdV equation. Given the
Hamiltonians
\[
\mathcal{H}_1 = \partial\,,\quad\quad
\mathcal{H}_2=\partial^3+\frac{2}{3} u \partial + \frac{1}{3} u_x
\]
one may form the recursion operator $ \mathcal{R}=\mathcal{H}_2\mathcal{H}_1^{-1}$
with which the symmetries take the form
\begin{equation}
u_{\tau_n} = \mathcal{R}^n u_x
\label{KdVflows}
\end{equation}
(for more details see, for example, \cite{O}). Such flows constitute the KdV hierarchy.

There has been much interest recently in the properties of the dispersionless
limits of integrable systems, these being examples of equations of hydrodynamic
type. As well as being interesting integrable systems in their own right,
they play an important role
in topological quantum field theories and the construction of higher-genus
Gromov-Witten invariants \cite{DZ1,DZ2}.
Under the scalings
\begin{eqnarray*}
t & \mapsto & \hbar^{-1} t \\
x & \mapsto & \hbar^{-1} x \\
u & \mapsto & u
\end{eqnarray*}
the KdV equation scales to
\[
u_t = u u_x + \hbar^2 u_{xxx}
\]
and under the classical limit $\hbar\rightarrow 0 $ one obtains the Monge equation
\begin{equation}
u_t = u u _x\,.
\label{Monge}
\end{equation}
This singular limit has a drastic effect on the solution space of the KdV
equation, for example soliton solutions to not survive, but many properties
do survive, notably the form of conservation laws and the existence of a bi-Hamiltonian
hierarchy. Thus a hierarchy of symmetries of the Monge equation may be
obtained by scaling the KdV hierarchy and taking the $\hbar\rightarrow 0 $ limit.
This results in the Monge hierarchy (also called in \cite{DZ2} the Riemann hierarchy)
\[
u_{\tau_n} = u^n u_x\,.
\]
However, the Monge equation has many more symmetries; given any suitably
differentible function $f(u)$ then
\begin{equation}
u_\tau = f(u) u_x
\label{generalMonge}
\end{equation}
is a symmetry, labelled not by a discrete integer, but by a $C^2(\mathbb{R})$-function.

The purpose of this paper is to study not the dispersionless limit but the reverse,
regarding the KdV equation as a deformation of the Monge equation and constructing,
term-by-term, the corresponding deformation of the general symmetry (\ref{generalMonge}).
In fact a slightly more general equation than the KdV equation will be studied, namely
the equation
\[
u_t = u u_x + \left(a(u) u_{xxx} + b(u) u_{xx} u_x + c(u) u_x^3 \right)
\]
where $a\,,b\,,c$ are arbitrary functions.
The motivation for this comes from the work of Eguchi et al. \cite{E}, Dubrovin and Zhang
\cite{DZ1,DZ2} and, in particular,
Lorenzoni \cite{L}. Their approaches utilize the bi-Hamiltonian structure of their systems.
No such assumption will be made here and, as a result, certain of Lorenzoni's results will
appear as a special case of the constructions that will appear in the subsequent sections.

\section{Notation and the form of the general problem}

Let $\D$ denote the linear vector space/ring of differential polynomials in
$\{u_1,u_2, \ldots\}$ where $u_1=u_x,u_2=u_{xx}$ etc., with coefficients in the space
$C^\infty( \mathbb{R})$ of infinitely differentible functions,
\[
\D = C^\infty( \mathbb{R})[u_1,u_2, \ldots]\,,
\]
so an arbitrary element $F\in\D$ takes the form
\[
F = \sum_{|I|<\infty} a_I(u) u_I
\]
where $a_I(u)\in C^\infty( \mathbb{R})$ and $I$ is a multi-index. To
each
monomial one may assign a scaling degree by counting the number of $x$-derivatives
it contains, so
\begin{eqnarray*}
deg(u_k) & = & k \,,\quad\quad k\geq 1\,,\\
deg(AB) & = & deg(A) + deg(B)\,, \quad\quad A\,,B \in \D
\end{eqnarray*}
and no degree is assigned to the coefficient functions (this is very
similar to the ideas used in the theory of normal forms \cite{KM}, but a
different grading is used there). With this, $\D$ decomposes into a
sum of sub-spaces/rings consisting of terms of the same degree
\[
\D = \bigoplus_{k=0}^\infty \D_k,.
\]
Thus, for example,
\[
\D_3 = span\{u_{xxx}, u_{xx} u_x,u_x^3\}\,.
\]
Clearly $dim\,\D_k={\rm~number~of~partitions~of~}k\,,$ denoted by $\#(k)$.
With this notation it is easy to describe the problem to be studied.

Consider the evolution equation
\begin{equation}
u_t = K_1
\label{basicexample2}
\end{equation}
where $K_1= u u_x \in \D_1$ and corresponding symmetry
\begin{equation}
u_\tau = Q_1
\label{baresymmetry}
\end{equation}
where $Q_1=f(u) u_x \in \D_1\,.$ Suppose the equation (\ref{basicexample2}) is
deformed by an arbitrary element $K_3\in \D_3\,:$
\begin{eqnarray*}
u_t & = & K[\hbar]\,,\\
& = & K_1 + \hbar^2 K_3\,.
\end{eqnarray*}
Can one construct a corresponding deformation of the symmetry
(\ref{baresymmetry})
\begin{eqnarray*}
u_\tau & = & Q[\hbar]\,,\\
& = & \sum_{n=0}^\infty \hbar^{2n} Q_{2n+1}
\end{eqnarray*}
with $Q_i\in\D_i$ and $Q_1$ as above? The deformation parameter $\hbar$ should be
regarded as a formal parameter, labelling the scaling dimensions of the terms it
precedes, rather than a small parameter. One may, at the end of all calculations,
set $\hbar=1\,.$

By definition, the symmetry $Q[\hbar]$ must satisfy the governing equation
\[
K[\hbar]^\prime Q[\hbar] - Q[\hbar]^\prime K[\hbar] = 0
\]
and equating powers of $\hbar$ gives the basic equation
\begin{equation}
K^\prime_1 Q_{2n+1} - Q_{2n+1}^\prime K_1= Q_{2n-1}^\prime K_3-K^\prime_3 Q_{2n-1}\,,\quad\quad
n\in \mathbb{N}\,.
\label{basicequation}
\end{equation}
By construction the equation for the lowest power $\hbar^0$ is automatically satisfied.
This equation (\ref{basicequation}) looks complicated but, since all the elements
lie in various spaces with fixed scaling dimension, it reduces to an over-determined linear
system.

\section{Integrable deformations of the Monge equation}

In this section equation (\ref{basicequation}) will be studied in more detail,
with
\begin{eqnarray*}
K_1 & = & u u_x \,, \\
K_3 & = & a(u) u_{xxx} + b(u) u_{xx} u_x + c(u) u_x^3\,, \quad\quad
a,b,c \in C^\infty( \mathbb{R})\,,\\
Q_1 & = & f(u) u_x\,,
\end{eqnarray*}
so
\begin{equation}
u_t = u u_x + \hbar^2 \left(a(u) u_{xxx} + b(u) u_{xx} u_x + c(u) u_x^3\right)\,.
\label{generalizedKdV}
\end{equation}
With these (\ref{basicequation})
becomes
\begin{equation}
\left[
\sum_{r=1}^{2n+1} [ (u u_1)_r - u u_{r+1}]
\frac{\partial~}{\partial u_r} - u_1
\right] Q_{2n+1} =
\left[
\begin{matrix}
(a^\prime u_3 + b^\prime u_2 u_1 + c^\prime u_1^3) + \\ ~~ \\
(a \partial^3 + b u_1 \partial^2 + b u_2 \partial + 3 u_1^2 \partial) - \\ ~~ \\
\sum_{r=0}^{2n-1} \partial^r ( au_3 + b u_2 u_1 + c u_1^3) \frac{\partial~}{\partial u_r}
\end{matrix}
\right] Q_{2n-1}
\end{equation}
or schematically as
\begin{equation}
\R(Q_{2n+1}) = \S(Q_{2n-1})
\label{operatorbasicequation}
\end{equation}
where
\begin{eqnarray*}
\R:\D_n \longrightarrow \D_{n+1} \,, \\
\S:\D_n \longrightarrow \D_{n+3} \,.
\end{eqnarray*}
One may regard this equation in a number of ways:

\begin{itemize}
\item{} as a first order quasi-linear equation in the independent
variables $u_1,u_2$ etc., and solve it using the method of characteristics.
However, one requires, for the purposes of this paper, the solution to lie in
$\D_{2n+1}$ which is hard to guarantee using this method;
\item{} as a problem in differential Galois theory - one requires the solution
to this linear equation with coefficients in a particular ring to lie in that
ring and not in some extension;
\item{} as an overdetermined linear system.
\end{itemize}
This third approach is the one that will be using here. This interpretation
holds because $\partial_u$ does not appear in the operator $\R$, so no derivatives
of the unknown coefficients appear. By introducing a basis for $\D_{2n+3}$ one obtains
$\#(2n+3)$ linear equations for $\#(2n+1)$ unknowns.

In the next section it will be shown that the rank of $\R$ (viewed as a matrix) is maximal,
and hence the solution to this linear problem, if it exits, is unique. Existence is
more problematic. While the form of the matrix $\R$ is easy to understand, the \lq source\rq~vector
$\S(Q_{2n-1})$ is complicated, depending on all lower order solutions, and so it would be
extremely difficult to compute, in general, the rank of the extended matrix.
By direct calculation (using Mathematica) explicit solutions up to $O(\hbar^6)$ have been
calculated. This suggest the conjecture that a solution exists for all $n$, so a formal symmetry of
equation (\ref{generalizedKdV}) of the form $\sum \hbar^{2n} Q_{2n+1} \in \D$ exists.

\section{Uniqueness}

One may to prove that the equation (\ref{operatorbasicequation}) has a unique solution
is to introduce an ordered basis for each subspace $\D_k$ of monomials $u_I$ -
the reverse lexiographic ordering -  so,
for example
\[
\D_5=span\{u_5,u_4u_1,u_3 u_2,u_3 u_1^2,u_2^2 u_1,u_2 u_1^3,u_1^5\}
\]
and this basis may be order
\[
u_5 \succ u_4u_1 \succ u_3 u_2 \succ u_3 u_1^2 \succ u_2^2 u_1 \succ u_2 u_1^3 \succ u_1^5\,.
\]
The symbol $max\{v\}$ will denote the highest basis vector in the expansion of $v$ in this
ordered basis.

\begin{lemma} In the reverse lexiographic ordered basis the operator
$\R:\D_n \longrightarrow D_{n+1}$ is lower-triangular.
\end{lemma}

\noindent It then follows that the equation $\R v=0$ has the unique solution $v=0$
and hence that the solution to (\ref{operatorbasicequation}), if it exists, is unique.

\medskip

\noindent{\bf Proof~} Let $u_I$ be a monomial in $\D_n\,,I$ being a multi-index. From the
explicit form of $\R$ it follows that
\[
max\{\R u_I\} = u_I u_1\,.
\]
Suppose that $u_I \succ u_J\,.$ Then
\[
max\{\R u_I\} = u_I u_1 \succ u_J u_1 = max\{\R u_J\}\,.
\]
Hence, in this basis, the matrix $\R$ is lower-triangular.

\bigskip

\noindent For example,
\[
\R [ g_1 u_{xxx} + g_2 u_{xx} u_x + g_3 u_x^3] =
3 g_1 u_{xxx} u_x + 3 g_1 u_{xx}^2 + 3 g_2 u_{xx} u_x^2 + 2 g_3 u_x^4
\]
so in the reverse lexiographic basis, $\R$ may be represented as the $3 \times 5$
matrix
\[
\R=\begin{pmatrix}
0 & 0 & 0 \\
3 & 0 & 0 \\
3 & 0 & 0 \\
0 & 3 & 0 \\
0 & 0 & 2
\end{pmatrix}\,,
\]
which clearly is of maximal rank.

\section{Existence}

Recall that the given data is $K_1\,,K_3$ and $Q_1\,.$ Writing
\[
Q_3=g_1 u_{xxx} + g_2 u_{xx} u_x + g_3 u_x^3
\]
and solving the over-determined linear system yields
\begin{eqnarray*}
g_1 & = & a f^\prime \,, \\
g_2 & = & b f^\prime + 2 a f^{\prime\prime}\,,\\
g_3 & = & c f^\prime + \frac{1}{2} b f^{\prime\prime} + \frac{1}{2} a f^{\prime\prime\prime}\,.
\end{eqnarray*}
Thus a solution exists. This gives an approximate symmetry, valid to $O(\hbar^2)\,.$
Recycling this solution gives
\begin{equation}
Q_5 = h_1\, u_5 + h_2 \, u_4 u_1 + h_3\, u_3 u_2 +
h_4 \,u_3 u_1^2 + h_5\, u_2^2 u_1 + h_6 \,u_2 u_1^3 + h_7 \,u_1^5
\label{fifthordersymmetry}
\end{equation}
where the $h_i$ are explicit functions of $a,b,c$ and $f$ and their derivatives.
These are given in the appendix.
The $O(\hbar^6)$ have also been calculated, again using Mathematica, but little
purpose is served by presenting them here - it suffices to say that they exist.
Thus for arbitrary functions $a,b,c$ a unique symmetry, labelled by a suitably
differentable function $f$ exists up to $O(\hbar^6)\,.$ The existence of solutions
up to this order of the over-determined linear system suggests the following:

\begin{conjecture} A unique formal symmetry to the generalized KdV equation
(\ref{generalizedKdV}) exists, labelled by a $C^\infty( \mathbb{R})$-function.
\end{conjecture}

\noindent If the conjecture is false, then this would raise the questions:

\medskip

\noindent{\bf Conjecture $\mathbf{1^\prime}$.~}{\it At what order does the above conjecture fail? Is there a way to determine this
order a priori?}

\medskip

\noindent Note that if the conjecture is true then one obtains an
entirely new integrable hierarchy, depending on three arbitrary functions and with
flows labelled by another arbitrary function. If it is false, then one may obtains
the notion of an
{\sl approximately} integrable system, which has an infinite number of conservation laws up to
some fixed order. Either outcome is of interest. In a particular case, Lorenzoni \cite{L} has
numerically observed elastic soliton scattering indicative of integrability. It would
be interesting to see if basic integrability results such as inverse scattering could
be modified (by, say, only including terms up to a given order) to include such approximate
integrable systems.


\subsection{Formal symmetries of the KdV equation}

In this approach, the KdV equation itself is recovered as the special case $a=1, b=c=0\,,$
and the symmetry takes the form
\begin{eqnarray*}
u_\tau & = & f(u) u_x + \hbar^2 \Bigg[ f^\prime u_{xx} + \frac{1}{2} u_x^2 f^{\prime\prime}
\Bigg]_x + \\
&  & ~\hbar^4 \Bigg[\frac{1}{8} f^{(5)} u_x^4 + \frac{11}{10} f^{(4)} u_x^2 u_{xx} +
\frac{9}{10} f^{(3)} u_{xx}^2 + \frac{6}{5} f^{(3)} u_x u_{xxx} + \frac{3}{5} f^{(2)} u_{xxxx}
\Bigg]_x + O(k^6)\,.
\end{eqnarray*}
(again the $O(\hbar^6)$ terms have been calculated explicitly, though not displayed here).
One may prove from the now simplifier version of (\ref{basicequation}) that for the
series to terminate the function $f$ must be a polynomial, say $f=\sum^N_{r=0} \alpha_r u^r\,,$
and one recovers a linear combination of the flows obtained via the recursion operator
(\ref{KdVflows}),
\[
u_{t_N} = \left( \sum_{r=0}^N \alpha_r \R^r \right) u_x\,.
\]
It would be of interest to see if the general flow generated from an arbitrary $f$
could be written as $u_\tau = F(\R ) u_x$ for some suitable $F\,.$

Much has been written about the symmetries of the KdV equation. However in such
approaches the order of the symmetry is fixed, and the lower-order terms are
determined from the higher ones. Here is approach is opposite - determine the
higher-order terms from lower ones. The symmetries obtained in this was will
turn out to be formal - infinite series. This approach is motivated,
as was mentioned in the introduction,
by ideas originating in topological quantum field theories, where the deformation
is known as a genus expansion, and from the deformation quantization
programme, where in both areas one constructs higher-order terms from lower-order ones.

\subsection{Local Hamiltonian systems}

An important subclass of systems in this class are those that are Hamiltonian with
respect to the local
Hamiltonian operator $ \mathcal{H}_1 = \partial\,,$ so the generalizes KdV
equation (\ref{generalizedKdV}) takes the from
\[
u_t = \mathcal{H}_1 \frac{\delta~}{\delta u} H\,
\]
for some Hamiltonian $H\,.$ This places the following restriction on the
arbitrary functions $a,b,c:$
\begin{eqnarray*}
a(u) & = & s(u)\,,\\
b(u) & = & 2 s^\prime(u) \,, \\
c(u) & = & \frac{1}{2} s^{\prime\prime}(u)
\end{eqnarray*}
where $s$ is an arbitrary function. With these (\ref{generalizedKdV}) may be written in
Hamiltonian form
\begin{eqnarray*}
u_t & = & u u_x + \hbar^2
\left[ s(u) u_{xxx} + 2 s^\prime (u) u_{xx} u_x + \frac{1}{2} s^{\prime\prime}(u)
u_x^3 \right]\,, \\ \\
 & = & \frac{d~}{dx}\frac{\delta~}{\delta u}
 \left[ \frac{u^3}{6} - \hbar^2 \frac{1}{2} s(u) u_x^2 \right]
\end{eqnarray*}
The formal symmetries are also Hamiltonian:
\[
u_\tau = \frac{d~}{dx}\frac{\delta~}{\delta u}
 \left[ f^{(-2)} - \hbar^2 \left\{ \frac{1}{2} s f^{\prime\prime}\right\} u_x^2 + \hbar^4 \left\{
 \frac{3}{10} s^2 f^{\prime\prime} u_{xx}^2 - \frac{1}{4!} \left[\frac{3}{2} (s^2)^{\prime\prime}
 f^{\prime\prime} + s^2 f^{(6)}\right] u_x^4 \right\} \right] + O(\hbar^6)\,
\]
(where $\partial^2 f^{(-2)}=f$). Again, the $ O(\hbar^6)$-terms have been calculated,
and are also Hamiltonian, which suggests the following:

\begin{conjecture} The formal system in Conjecture 1 is Hamiltonian to all orders.
\end{conjecture}

\noindent This system has been extensively studied, for $f(u)=u^n\,,$ by Lorenzoni \cite{L}.
He showed that the systems is, up to $O(\hbar^4)$, bi-Hamiltonian,
with the second Hamiltonian structure being a deformation of the
second Hamiltonian structure of the Monge/Riemann hierarchy. The existence of such
terms in this deformation is controlled by a certain cohomology group.

\section{Trivial and quasi-trivial Miura transformations}

In \cite{DZ2} transformation of the form,
\[
u \mapsto v = u + \sum_{k=1}^\infty \hbar^k {\widetilde F}_k(u;u_x,\ldots,u^{(n_k)}),
\]
where ${\widetilde F}_k$ are quasi-homogeneous rational functions, were considered together with the
corresponding action on bi-Hamiltonian pencils. Such transformations were
called quasi-Miura transformations. In particular the notion
of trivial and quasi-trivial transformations were given. In the context of this
paper, where local bi-Hamiltonian structures are not consider, an evolution
equation $u_t=u u_x + \sum_r \hbar^r K_r[u]$ will be said to be quasi-trivial if it
transforms under a quasi-Miura transformation to
the Monge equation $v_t=v v_x$, and trivial if the further condition that the
functions $F_k$ are polynomial is satisfied.

With the ansatz
\[
v=u+\sum_{r=1}^\infty \frac{\hbar^{2r}}{u_x^{4r-2}}\, F_r[u,u_1,\ldots,u_{6r-2}]\,,\quad\quad F_r\in {\mathcal{D}}_{6r-2}
\]
one may easily obtain recursion relations for the $F_r$ which take the form
\[
\widetilde{\R}(F_n) = {\rm function~of~}F_1\,,\ldots\,,F_{n-1}
\]
where
\[
\widetilde{\R}:\D_n \longrightarrow \D_{n+2}
\]
is given by
\[
\widetilde{\R} = u_1 \R +(2-4n) u_1^2
\]
with $\R$ being the previously introduced recursion operator. It follows immediately
from a repetition of the argument in section 4 this that the quasi-Miura transformation, if it exists, is unique, since the
operator $\widetilde{\R}$ is lower triangular in the
reverse lexiographic ordered basis.

\begin{lemma} The generalised KdV equation (\ref{generalizedKdV}) is, up to $O(\hbar^4)$,
quasi-trivial.
\end{lemma}

\noindent This may be proved by direct calculation. The quasi-Miura transformation being
\[
v=u + \frac{\hbar^2}{u_x^2}
\left( \frac{1}{2} a(u) (u_{xxx} u_x- u_{xx}^2) + \frac{1}{2} b(u) u_{xx} u_x^2 + c(u) u_x^4\right)
+\frac{\hbar^4}{u_x^6}  f(u,u_x\,,\ldots\,,u_{10x} )\,,
\]
where $f\in {\mathcal{D}}_{10}$ has been explicitly calculated, though not displayed here.
In \cite{DZ2} it was shown that the KdV equation is quasi-trivial to all orders, this
proof using the existing machinery for the KdV equation. Conjecturally the generalized KdV equation
studied here will be quasi-trivial to all orders, though to prove it would involve the
development of a lot of associated machinery.

\medskip

One immediate question is whether the quasi-trivial equation (\ref{generalizedKdV}) can,
for suitable choice of functions $a\,,b\,,c$ be trivial. From the above form of the
quasi-Miura transformation it follows that at $O(\hbar^2)$ one requires $a(u)=0$ and at
$O(\hbar^4)$ one requires, in addition, $b(u)=0\,.$ After this no new constraints appear,
at least to $O(\hbar^{10})\,,$ the explicit quasi-Miura map being
\begin{eqnarray*}
v&=&u-\hbar^2  c \, u_1^2 + \hbar^4\left(
\frac{4}{3}\,c\,c'\,u_1^4 + 2\,c^2\,u_1^2\,u_2\right)\\
&&+\hbar^6 \left(
\frac{-28\,c\,{c'}^2\,{u_1}^6}{15} -
  \frac{14\,{c}^2\,c''\,{u_1}^6}{15} -
  \frac{26\,{c}^2\,c'\,{u_1}^4\,u_2}
   {3} - 4\,{c}^3\,{u_1}^2\,{u_2}^2 -
  \frac{4\,{c}^3\,{u_1}^3\,u_3}{3}
\right)\\
&&+\hbar^8 \left(
\frac{836\,c\,{c'}^3\,{u_1}^8}{315} +
  \frac{1268\,{c}^2\,c'\,c''\,{u_1}^8}
   {315} + \frac{16\,{c}^3\,c^{(3)}\,{u_1}^8}
   {35} + \frac{1138\,{c}^2\,{c'}^2\,{u_1}^6\,
     u_2}{45}\right.\\
     &&\left. + \frac{128\,{c}^3\,c''\,
     {u_1}^6\,u_2}{15} +
  \frac{112\,{c}^3\,c'\,{u_1}^4\,
     {u_2}^2}{3} + 8\,{c}^4\,{u_1}^2\,
   {u_2}^3 + 8\,{c}^3\,c'\,{u_1}^5\,
   u_3 + 8\,{c}^4\,{u_1}^3\,u_2\,
   u_3 + \right.\\
     &&\left.\frac{2\,{c}^4\,{u_1}^4\,
     u_4}{3}
\right)+O(\hbar^{10})\,.
\end{eqnarray*}
Thus the first order, third degree evolution equation
\[
u_t = u u_x + \hbar^2 c(u) u_x^3
\]
is trivial to $O(\hbar^8)$ and conjecturally to all orders. Further
properties of this system are outlined in the next section.

\section{First order deformation of arbitrary degree}

Exact results to all orders may be obtained in the above special case $a=b=0, c(u)=\alpha(u)\,.$
The basic equation (\ref{generalizedKdV}) simplifies to
\begin{equation}
u_t = u u_x + \hbar^2 \alpha(u) u_x^3\,,\quad\quad \alpha \in C^\infty( \mathbb{R})
\label{fotd}
\end{equation}
and the governing equation (\ref{basicequation}) simplifies drastically. With the
ansatz
\[
Q_{2n+1} = \beta_n(u) u_x^{2n+1}
\]
one obtains the recursion relation
\[
n \beta_n = (1-n) \alpha^\prime \beta_{n-1} + \alpha \beta^\prime_{n-1}\,, \quad\quad n\geq 1\,,
\]
and hence, since $\beta=\beta_0$ is arbitrary, the solution
\[
\beta_n = \frac{1}{n!} \alpha^n \beta^{(n)}\,.
\]
Thus the general symmetry of the first order, third degree equation
(\ref{fotd}) is
\[
u_{\tau_\beta} = \sum_{n=0}^\infty \hbar^{2n} \frac{\alpha^n \beta^{(n)}}{n!} u_x^{2n+1}\,.
\]
Thus the space of commuting flows for equation (\ref{fotd}) is labelled by an arbitrary
function $\beta\in C^\infty( \mathbb{R})\,.$ This series truncates if and only if $\beta$
is a polynomial. This symmetry is, within the constraints of this paper, unique. It can
easily be checked that all the symmetries commute amongst themselves, i.e.
\[
(u_{\tau_\beta})_{\tau_\gamma}= (u_{\tau_\gamma})_{\tau_\beta}
\]
for arbitrary $C^\infty$-functions $\alpha,\beta,\gamma\,.$

The equation (\ref{fotd}) is Hamiltonian with respect to the non-local
Hamiltonian operator
\[
\mathcal{H} = u_x \partial^{-1} u_x\,,
\]
so
\[
u_t = \left(\mathcal{H}\frac{\delta~}{\delta u} \right)
\int ( u - \hbar^2 \alpha(u) u_x^2) \, dx\,.
\]
The symmetry itself is also Hamiltonian,
\[
u_{\tau_\beta} =\mathcal{H}\frac{\delta H_\beta}{\delta u}
\]
where the Hamiltonian is
\[
H_\beta = \int \sum_{r=0}^\infty \frac{\hbar^{2r}}{(1-2r)} \alpha^r \beta^{(r)} u_x^{2r} \, dx
\]
and these are all conserved with respect to all the symmetries, as one would expect
for an integrable system. Owing to the arbitrariness in the functions $\alpha$ and $\beta$
is unlikely that (\ref{fotd}) will possess the Painlev\'e property.

\section{Conclusions}

The central philosophy behind this paper is that the dispersionless limits of
integrable systems should not been seen as some limit, but as the kernel of the
dispersive hierarchy itself, encapsulating its integrability. To describe the
full dispersive hierarchy one just has to specify a suitable \lq deformation vector\rq~
in some space of possible first-order deformations, from which the dispersive hierarchy,
if it exists,
may be uniquely reconstructed. In this paper this vector is $K_3$, living in the
space of possible first-order deformations $\D_3\,.$ If the above conjecture is true, then
this specifies the hierarchy at all orders. For equations of hydrodynamic type constructed
from semi-simple Frobenius manifolds this choice is automatic \cite{DZ1,DZ2} - elliptic
Gromov-Witten invariants may be constructed from the genus-zero data. See also \cite{E},
where examples are given where higher-order deformations do not exist.

As an multi-dimensional example consider the system
\[
\mathbf{U}_t = \mathbf{U} \circ \mathbf{U}_x\,,
\]
where $\circ$ is the product is some $N$-dimensional, commutative $N$-dimensional algebra.
With ceratin simple additional conditions on this algebra one may show that a
hierarchy exists if and only if the algebra is a Jordan algebra \cite{S}. It then may
be shown that the deformed system
\[
\mathbf{U}_t = \mathbf{U} \circ \mathbf{U}_x + \hbar^2 \mathbf{U}_{xxx}
\]
(corresponding to the deformation vector
\[
\mathbf{U}_{xxx}\in\D_3^N=span\{u^i_{xxx},u^i_{xx} u^j_x,u^i_xu^j_xu^k_x\,,i,j,k=1,\ldots,N\}
\]
where $\mathbf{U}=u^i e_i$ and the $e_i$ form a basis for the algebra) defines a full
dispersive hierarchy with no extra conditions required. It would be interesting to repeat these
calculations for a more general vector in $\D^N_3$ - the results in this paper being for
$N=1$ only.

It may also be of interest to generalize the construction, in particular the
quasi-triviality property, to include $K_5$-terms in
(\ref{basicequation}),
\[
u_t=K_1 + \hbar^2 K_3 + \hbar^4 K_5\,.
\]
A number of known systems fall into this class (with $K_1=u^2 u_x$), including the higher-flow of the
$KdV$ equation, $KdV_5\,,$
the Sawada/Kotera equation, the Caudrey/Dodd/Gibbons/Kaup equation and the Kaup/Kupershmidt
equation. This paper suggests a possible classification of such fifth-order systems.
Also, little mention has been made of bi-Hamiltonian structures in this paper. Such structures may
be generated from the known bi-Hamiltonian pencil of the Monge equation
\[
\{u(x),u(y)\}_\lambda = (u-\lambda) \delta^\prime(x-y) + \frac{1}{2} u_x \delta(x-y)
\]
via the appropriate quasi-Miura transformation. These, however, are unlikely to truncate.

\medskip

The results outlined here have been computational; to obtain rigorous results at all orders
in the expansion new techniques will have to be developed.
In particular, a better understanding of the
over-determined linear systems from which results of this paper were obtained will be required.

\appendix
\renewcommand{\thesection}{}
\section{The $O(\hbar^4)$ order terms}

The explicit form of the $h_i$ in equation (\ref{fifthordersymmetry})
are:
\begin{eqnarray*}
h_1&=&(3 a^2 f^{\prime\prime})/5\,,\\
h_2&=&a b f^{\prime\prime} +
  (8 a a^{\prime} f^{\prime\prime})/5 +
  (9 a^2 f^{(3)})/5\,,\\
h_3&=&2 a b f^{\prime\prime} +
  2 a a^{\prime} f^{\prime\prime} +
  3 a^2 f^{(3)}\,,\\
h_4&=&(2 b^2 f^{\prime\prime})/5 +
  (11 a c f^{\prime\prime})/5 +
  b a^{\prime} f^{\prime\prime} +
  (13 a b^{\prime} f^{\prime\prime})/10 +
  (7 a f^{\prime\prime} a^{\prime\prime})/5 +\\&&
  (5 a b f^{(3)})/2 +
  (37 a a^{\prime} f^{(3)})/10 +
  (23 a^2 f^{(4)})/10\,,\\
h_5&=&(8 b^2 f^{\prime\prime} +
    5 b (a^{\prime} f^{\prime\prime} +
       7 a f^{(3)}) +
    a (14 c f^{\prime\prime} +
       26 b^{\prime} f^{\prime\prime} +
       13 f^{\prime\prime} a^{\prime\prime} +
       44 a^{\prime} f^{(3)} +
       31 a f^{(4)}))/10\,,\\
h_6&=&(9 b^2 f^{(3)} +
    b (18 c f^{\prime\prime} +
       9 b^{\prime} f^{\prime\prime} +
       4 f^{\prime\prime} a^{\prime\prime} +
       15 a^{\prime} f^{(3)} +
       25 a f^{(4)}) +\\&&
    2 (4 c a^{\prime} f^{\prime\prime} +
       10 a c^{\prime} f^{\prime\prime} +
       7 a f^{\prime\prime} b^{\prime\prime} +
       16 a c f^{(3)} +
       14 a b^{\prime} f^{(3)} +
       12 a a^{\prime\prime} f^{(3)} +\\&&
       2 a f^{\prime\prime} a^{(3)} +
       18 a a^{\prime} f^{(4)} +
       8 a^2 f^{(5)}))/10\,,\\
h_7&=&(4 c^2 f^{\prime\prime} +
    b^2 f^{(4)} +
    2 c (b^{\prime} f^{\prime\prime} +
       2 b f^{(3)} +
       a^{\prime} f^{(3)} +
       2 a f^{(4)}) +\\&&
    b (2 c^{\prime} f^{\prime\prime} +
       f^{\prime\prime} b^{\prime\prime} +
       2 b^{\prime} f^{(3)} +
       a^{\prime\prime} f^{(3)} +
       2 a^{\prime} f^{(4)} +
       2 a f^{(5)}) +\\&&
    a (2 f^{\prime\prime} c^{\prime\prime} +
       4 c^{\prime} f^{(3)} +
       3 b^{\prime\prime} f^{(3)} +
       f^{(3)} a^{(3)} +
       f^{\prime\prime} b^{(3)} +
       3 b^{\prime} f^{(4)} +
       3 a^{\prime\prime} f^{(4)} +
       3 a^{\prime} f^{(5)} +
       a  f^{(6)} ))/8
\end{eqnarray*}
The fifteen coefficients in the next-order term have been calculated, but
are not presented here.

\end{document}